\documentclass[]{aa}
\usepackage{graphicx}
\usepackage[applemac]{inputenc}
\usepackage{txfonts}

\begin{document}

\title{New constraints on the primordial black hole number density\\
from Galactic $\gamma$-ray astronomy}

\titlerunning{New constraints on the primordial black hole number density}

\author{R. Lehoucq\inst{1} \and M. Cassé\inst{1,2} \and J.-M. Casandjian\inst{1} \and I. Grenier\inst{1}}

\offprints{R. Lehoucq}
\institute{
Laboratoire AIM, CEA-Irfu/CNRS/Universit\'e Paris Diderot, Service d'astrophysique, CEA-Saclay, 91191 Gif-sur-Yvette, France
\and
Institut d'Astrophysique, 98 bis boulevard Arago, 75014 Paris, France
}

\date{Received date; accepted date}

\abstract
{Primordial black holes are unique probes of cosmology, general relativity, quantum gravity and non standard particle physics. They open a new window on the very small scales in the early Universe and also can be considered as the ultimate particle accelerator in their last (explosive) moments since they are supposed to reach, very briefly, the Planck temperature.  
}
{Upper limits on the primordial black hole number density of mass $M_{\star} = 5\,10^{14}$~g, the Hawking mass (born in the big-bang terminating their life presently), is determined comparing their predicted cumulative $\gamma$-ray emission, galaxy-wise, to the one observed by the EGRET satellite, once corrected for non thermal $\gamma$-ray background emission induced by cosmic ray protons and electrons interacting with light and matter in the Milky Way.
}
{A model with free gas emissivities is used to map the Galaxy in the 100~MeV photon range, where the peak of the primordial black hole emission is expected. The best gas emissivities and additional model parameters are obtained by fitting the EGRET data and are used to derive the maximum emission of the primordial black hole of the Hawking mass, assuming that they are distributed like the dark matter in the Galactic halo.
}
{The bounds we obtain, depending on the dark matter distribution, extrapolated to the whole Universe ($\Omega_{PBH}(M_{\star}) = 2.4\,10^{-10}$ to $2.6\,10^{-9}$ are more stringent than the previous ones derived from extragalactic $\gamma$-ray background and antiprotons fluxes, though less model dependent and based on more robust data.
}
{These new limits have interesting consequences on the theory of the formation of small structures in the Universe, since they are the only constraint on very small scale density fluctuations left by inflation. Significant improvements by data gathered by the FERMI $\gamma$-ray satellite are expected in the near future. The interest of a generalisation of this work beyond the standard particle model and in extradimensions is briefly alluded.
}

\keywords{Black hole physics; (Cosmology:) dark matter; Galaxy: center; Gamma rays: observations; Gamma rays: theory}

\maketitle

\section{Introduction}
\label{sec:introduction}

Black holes are classically defined by the property that things, including light, can go in but can't go out. However, this mere definition should be modified if quantum effects are taken into account, since they allow particle emission, with potentially observable effects (Hawking \cite{hawking75}, Page \& Hawking \cite{page76}). The Hawking radiation is in fact compatible with the well known result of general relativity that nothing can escape from inside the black hole horizon. Involving quantum fluctuations of the vacuum it can be conceived as the creation of virtual particle--antiparticle pairs that are disrupted by tidal forces in the vicinity of the horizon. The particle (antiparticle) with positive total energy escapes to infinity while the antiparticle (particle) with negative total energy falls into the black hole. This hypothetical process points to profound connections between thermodynamics, quantum mechanics and general relativity. Unfortunately, astrophysical black holes, either stellar or supermassive, radiate so weakly that they remain practically black.

On the other hand, beyond the stellar black holes and supermassive ones sitting at the centre of galaxies (Reid \cite{reid08}), well motivated observationally, a very much lighter category of primordial origin has been postulated (Carr \cite{carr74}). The formation of black holes in the early Universe (primordial black holes, hereafter PBHs) is regarded as an inescapable consequence of cosmological models, specially inflationary onces (for a review see Carr \cite{carr05a}). PBHs can thus serve as astrophysical probes of unification and their discovery, even if only minuscule ones are expected to show up through their radiation, would be a landmark on the way to quantum gravity.

Formed by pure gravity, PBHs require no special extension of the standard model of particle physics and are predicted on quite general grounds to be created in the very early Universe, though the mechanism of their formation remains conjectural. If they exist, they are expected to be much more emissive than the massive ones. Those tiny black holes have not yet been discovered in spite of considerable efforts (see Barrau et al. \cite{barrau03} and Carr \cite{carr05b} for reviews), but they are still searched for since their discovery would confirm the Hawking theoretical construction. Even an upper limit on their number density is of interest since, being the first gravitational collapsed objects to form, they could give precious information on the early Universe (Carr \cite{carr05a, carr05b}).

Particle emission from black holes has been predicted in a 4D spacetime in the framework of semi-classical quantum gravity (Hawking \cite{hawking75}, Page \cite{page76, page77}, Carr \& McGibbon \cite{carr98}). It has been recently extended to extradimensions in the brane scenario (Argyres et al. \cite{argyres98}, Emparan et al. \cite{emparan00}, Emparan \& Reall \cite{emparan08}, Kanti \cite{kanti04}). We here focus on the 4D case, postponing the $(4+n)$D case to a forthcoming paper. We only mention an interesting coincidence between 4D and 10 or 11D black holes.

A first constraint has been imposed on the mean mass density of $M = M_{\star}$ primordial black holes (in units of the critical density, $\Omega_{PBH}(M_{\star})$) through the extragalactic $\gamma$-ray background integrated both over the PBH mass spectrum and the lifetime of the Universe (Page \& Hawking \cite{pagehawking76}, McGibbon \& Carr \cite{mcgibbon91}, Carr \& McGibbon \cite{carr98}, improved by Barrau et al. \cite{barrau03}). This limit, however, depends on poorly  known quantities. First, the initial mass function of primordial black holes is purely conjectural because the conditions in the early Universe before decoupling are poorly constrained and large density contrasts could have arisen (Blais et al. \cite{blais03}).  Second, the true extragalactic gamma-ray background is difficult to extract both observationnally, because it could be contaminated by very local inverse-Compton emission and instrumental backgrounds, and theoretically, due to unresolved AGN contribution. At last, a similarity between the intensities of the PBH galactic and extragalactic gamma-ray fluxes in this energy range has been advocated by Wright (\cite{wright96}), based on its analysis of the EGRET observations, which we confirm in this work. It can be readily be explained noting that the dark matter column densities toward the galactic centre ($\approx 0.3$~g/cm$^3$) is, by chance, of the same ordrer of magnitude than the extragalactic one integrated over a large part of the observable universe ($\approx 0.2$~g/cm$^3$). Thus it is reasonable to search for clues of primordial black holes in the diffuse galactic flux rather than in the extragalactic one, since it does not involve the redshift.

A second constraint has been also obtained through the observation of antiprotons in the galactic cosmic rays (Maurin et al. \cite{maurin01}, Barrau et al. \cite{barrau02b}, and references therein). However, the limitation of his method is that the antiproton propagation is subject to uncertainties in their diffusion coefficient and the scale height of their confinement region. 

Here we search for the signature of PBHs in the Galactic $\gamma$-ray background, more intense and better known than the extragalactic one, taking benefit of the expected large concentration of PBHs in the central region of the Galaxy. More precisely, we assume that PBHs are broadly distributed like dark matter in the halo of our Galaxy. Admittedly, more sophisticated clustering as proposed by Chisholm (\cite{chisholm 06}) should be analyzed in the future. This method requires a careful modelling of the Galactic background, but it is free of the ambiguities of the initial PBH mass spectrum since it implies only black holes radiating now (i.e. of mass $M_{\star} = 5\,10^{14}$~g for a 4D universe). It is not plagued either by the large uncertainty in the unresolved source contribution to the extragalactic background.

\section{Hypothesis and ingredients}
\label{sec:hypothesis}

Our aim is to calculate the Galactic $\gamma$-ray emission due to putative PBHs assumed to be distributed as dark matter. We thus focus on those born during the very early phases of the Universe, 13.7 Gyr ago, and dying presently. Their initial mass and temperature are estimated to $M_{\star} = 5\,10^{14}$~g and $T_{\star} = 20$~MeV (Carr \& McGibbon \cite{carr98}, MacGibbon \& Webber \cite{mcgibbon90}). Because a black hole will emit all particle species with rest mass less than or equivalent to the black hole temperature, the emission is insensitive to the quark and gluon production which occurs above 200-300~MeV. The only species of interest in our case are light particles (neutrinos, gravitons, photons and electron-positron pairs) if low mass non standard particles are excluded. Their relative proportions (branching ratios) are rather well established. 

The main hypotheses of this work are the following :
\begin{enumerate}
  \item spacetime is 4D;
  \item PBHs form through a cosmological scenario;
  \item most PBHs are presently neutral and non rotating;
  \item being part of the dark matter, PBHs are distributed alike.
\end{enumerate}

Like previous authors, we adopt points 1, 2 and 3 without discussion, point 4 is reasonable and generally assumed (MacGibbon \& Carr \cite{mcgibbon91}, Barrau \cite{barrau02}). Two ingredients are crucial to perform the calculation: the $\gamma$-ray emission spectrum of an individual black hole with mass $M_{\star}$ and the Galactic distribution of PBHs. The first one, relying on detailed calculations of McGibbon \& Webber (\cite{mcgibbon90}) is rather well understood. The second one lies on a more shaky ground, since the distribution of PBHs in the Galaxy inherits all the uncertainties of the dark matter distribution. We will show, however, that this uncertainty does not invalidate our analysis.

\subsection{PBH emission spectrum}

\begin{figure}
\begin{center}
\includegraphics[width=8cm]{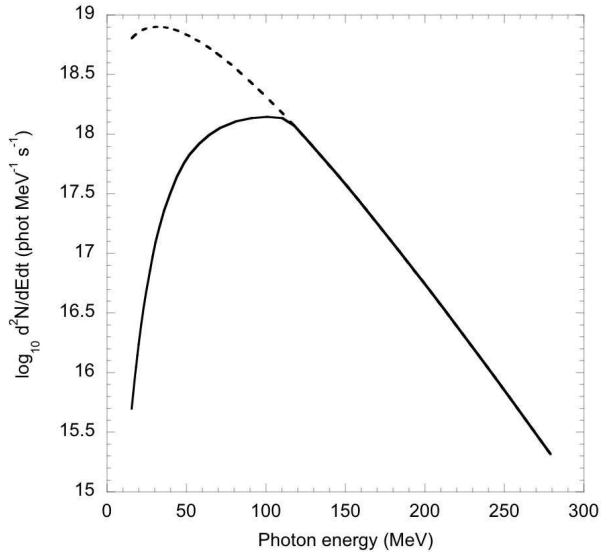}
\caption{Instantaneous $\gamma$-ray photon emission rate from a $kT = 20$~MeV black hole (McGibbon \& Webber \cite{mcgibbon90}) compared to the black-body emission at the same temperature (dashed line).}
\label{fig:spectrePBH}
\end{center}
\end{figure}

The spectrum of the Hawking radiation emanating from a black hole with temperature $T_\mathrm{H}$ is not purely thermal since it has not an exact blackbody profile. The correcting factor, also called greybody factor, depends on the energy of the emitted particles and strongly on their spin (high spin being defavoured). It significantly modifies the emitted spectrum below $k T_\mathrm{H}$. This distortion can be attributed to the fact that any particle emitted near the black hole horizon has to traverse a strong gravitational background before reaching the observer at infinity, at variance with what happens with a black body in a flat spacetime (Page \cite{page76}). Consequently, the flux peaks at higher energy (around $5\,k T_\mathrm{H}$) than for a pure blackbody at same temperature (which flux is maximum at $1.59~kT$); see Figure \ref{fig:spectrePBH}. The spin dependence of the greybody factor determines the relative emissivities (branching ratios) in different particle types from the black hole.

Since the typical temperature of PBHs born in the early Universe and that end its life at present time is about 20~MeV, a distinctive signature of quantum black holes would be a quasi-planckian spectrum at unexpectedly high energy, peaking at about 100 MeV. This makes them particularly suitable to observation by $\gamma$-ray satellites. We thus reassess the question of limiting their maximal number in the light of EGRET observations of the diffuse $\gamma$-ray emission of the Galaxy complementary to previous attempts relying on the extragalactic $\gamma$-ray background (Page \& Hawking \cite{pagehawking76}, Halzen et al. \cite{halzen91}, Carr \& MacGibbon \cite{carr98}, Kim et al. \cite{kim99}, Barrau et al. \cite{barrau03}) and antimatter in cosmic rays (Boudoul \& Barrau \cite{boudoul03}).

\subsection{Dark matter density distribution}

It is reasonable to assume that PBHs are a minor part of dark matter and that, as such, they follow its calculated distribution. Thus, the expected $\gamma$-ray flux is proportional to the dark matter density which is quite different from the scenario based on neutralino annihilation where the $\gamma$-ray flux varies quadratically with dark matter density. The dark matter distribution, deduced from numerical simulations (Navarro, Frenk \& White \cite{navarro97}, Moore et al. \cite{moore99}) is conjectural and quite uncertain. We use the most advertised dark matter distributions in the literature. For the sake of simplicity, we do not discuss clumping in this article and consider only smoothed distributions. The inner Galaxy ($\approx 100$~pc from the center), where the dark matter density (and hence presumably the PBH density) is larger, is the main place for dark matter indirect searches through $\gamma$-ray signatures. Indeed, since the total mass in the inner Galaxy is dominated by baryons, the dark matter distribution is likely to have been influenced by the baryonic potential (Bertone \& Merrit \cite{bertone05}). Adiabatic compression models including the effect of the baryonic gas  significantly increase the dark matter density in the central region of the Milky Way, and as such, lead to stronger constraints on the PBH number density. More realistic than the standard model, they are unfortunately still quite uncertain. The halo models with adiabatic compression have been discussed and put under a useful form by Mambrini et al. (\cite{mambrini06}, see their table 1), from which we borrow the results. The Navarro-Frenk-White (hereafter NFW) distribution diverges in the Galactic centre direction. To cure this divergence we use the prescription from Delahaye et al. (\cite{delahaye08}).

\begin{figure}
\begin{center}
\includegraphics[width=8cm]{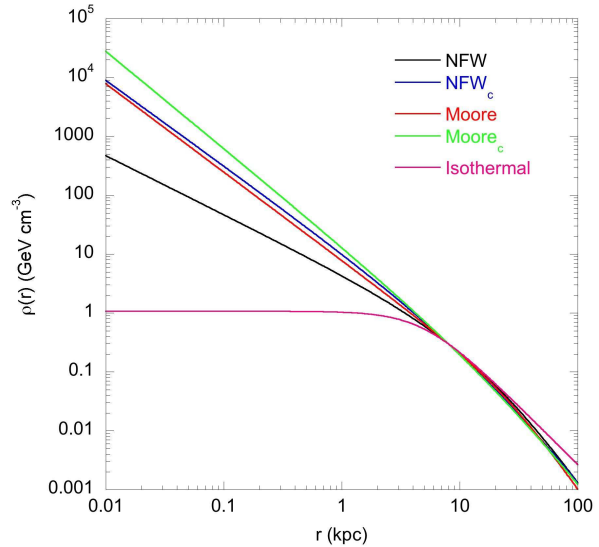}
\caption{Dark matter density profiles following NFW (\cite{navarro97}), Moore et al. (\cite{moore99}) without and with adiabatic compression (NFW$_\mathrm{c}$ and Moore$_\mathrm{c}$ respectively, Mambrini et al. \cite{mambrini06}). An isothermal profile is also added for comparison. All these distributions are normalized to 0.3~GeV/cm$^3$ in the solar neighbourhood.}
\label{fig:distributionPBH}
\end{center}
\end{figure}

Note that a 4 millions solar masses supermassive black hole sits at the centre of the Milky Way. Adiabatic accretion of dark matter onto this object could lead to a very high density of dark matter (Bertone \& Merrit \cite{bertone05}). An initial dark matter density profile $\rho(r) \propto r^{-\alpha}$ would lead to a final density profile with a slope $\alpha' = (9 - 2 \alpha/(4 - \alpha)$ so sharp that it is equivalent to a point source for EGRET and Fermi satellites.

\section{Modelling the $\gamma$-ray emission from the interstellar medium and PBHs in the Milky Way}
\label{sec:fit}

Our aim is to estimate the maximum $\gamma$-ray flux from PBH emission in the Galaxy that is allowed by observations on top of the cosmic-ray induced $\gamma$-rays of interstellar origin. More precisely, we compare the total intensity expected from the various PBH distributions and known interstellar components to the whole-sky survey data obtained by the Energetic Gamma-Ray Experiment Telescope (EGRET). The latter, which operated onboard the Compton-Gamma Ray Observatory from April 1991 to May 2000, detected photons in the 20 MeV to 30 GeV range. The observation program made use of the large instrumental field of view to cover the entire sky and to complete in-depth studies of specific regions. EGRET is well suited for this study since the peak energy of PBH emission ($E\approx 100$~MeV) is inside the range of high effective area of the detector (Thompson \cite{thompson93}). Moreover, EGRET repeatedly observed the Galactic centre and Galactic plane regions which thus benefit from long exposures and limited off-axis aberrations.

The Galactic emission is produced by the interaction of energetic cosmic-ray electrons and protons with interstellar nucleons and photons. The decay of neutral pions produced in hadron collisions, the inverse-Compton scattering of the interstellar radiation field by electrons and their bremsstrahlung emission in the interstellar gas are the main contributors to the Galactic emission. If energetic cosmic rays penetrate uniformly all gas phases, the $\gamma$-ray intensity in each direction of the sky can be modelled by a linear combination of gas column-densities, an inverse-Compton intensity map ($I_{IC}$), and an isotropic intensity ($I_{iso}$) which accounts for the very local inverse-Compton emission, the extragalactic background, and the residual instrumental background. To account for the non-uniform cosmic-ray flux in the Galaxy, the gas column densities are distributed within galactocentric rings. The all-sky Leiden-Argentina-Bonn (LAB) composite survey (Kalberla et al. \cite{kalberla05}) was used to obtain the atomic hydrogen $N_{HI}$ column densities assuming a constant spin temperature of 125~K. The molecular hydrogen column densities were inferred from the velocity integrated CO intensity  $W(CO)$ obtained from the Center for Astrophysics compilation (Dame et al. \cite{dame01}). Following Strong and Mattox (\cite{strong96}), we assumed a uniform $X_{CO} = N(H_{2})/W(CO)$ conversion ratio with galactocentric distance to reduce the number of free parameters in the model. The gas column densities within 10$^{\circ}$ of the Galactic centre and anticentre were interpolated from the regions just outside these boundaries, then normalized to match the total emission observed along the line-of-sight. We have also included within the solar-circle ring the large column-densities $N_{Hdark}$ of ``dark" gas associated with cold and anomalous dust at the transition between the atomic and molecular phases in the nearby clouds (Grenier et al. \cite{grenier05}). To derive the spatial distribution and intensity of the inverse-Compton radiation, we have used the GALPROP model for cosmic-ray propagation developed by Strong et al. (\cite{strong07}), more precisely the version 50p of this model with an electron source spectrum similar to that measured near Earth. This version assumes a cosmic-ray source distribution matching the radial distribution of Galactic pulsars from Johnston (\cite{johnston94}). Maps of PBH emission, $I_{PBH}(l,b)$, were constructed as a function of energy using the spectrum displayed in Figure \ref{fig:spectrePBH} for the various radial profiles of dark matter densities given in Figure \ref{fig:distributionPBH}. Our model also includes the flux and spatial distribution of all point sources that were detected above 6$\sigma$ in the revised EGRET catalogue (Casandjian \& Grenier \cite{casandjian08}) .

The predicted count rates per bin in a model map are therefore calculated as:
$$
\begin{array}{l}
N_{pred}(l,b) =  \epsilon(l,b)\, \big( q_{IC} I_{IC}(l,b) + I_{iso} + q_{dark} N_{Hdark}(l,b) \\
+ \sum_{i=rings} q_{HI,i} [N_{HI}(r_i,l,b) + 2X_{CO}\,W_{CO}(r_i,l,b)] \\
+ q_{PBH} I_{PBH}(l,b) \big) + \sum_{j=sources} \epsilon(l_j,b_j)\, f_{j}\, PSF(l_j,b_j)
\label{eq:eqRing}
\end{array}
$$

Here $f_{j}$ and $\epsilon(l,b)$ represent source fluxes and the EGRET exposure as a function of Galactic coordinates, respectively. The product of the diffuse maps with exposure were convolved with the EGRET point-spread function (PSF) for an input $E^{-2.1}$ spectrum before adding the source maps. The EGRET data, exposure, and instrument response functions were downloaded from the CGRO Science Support Center. The $q$ parameters (gas emissivities or relative contributions of different radiation components), the isotropic intensity, and the $X_{CO}$ factor were fitted to the EGRET photon maps in six energy bands, namely 70-100 MeV, 100-150 MeV, 150-300 MeV, 300-500 MeV, 500-1000 MeV, and 1000-2000 MeV. We used a 2-dimensional maximum-likelihood method with Poisson statistics using maps in $0.5^{\circ}\times 0.5^{\circ}$ bins. The likelihood ($L$) is calculated as the product, for all pixels at $10^{\circ} \leq |l| \leq 170^{\circ}$ and all latitudes, of the Poisson probabilities of observing $N_{obs}(l,b)$ photons in a pixel where the model predicts $N_{pred}(l,b)$.

\begin{figure}
\begin{center}
\includegraphics[width=8cm]{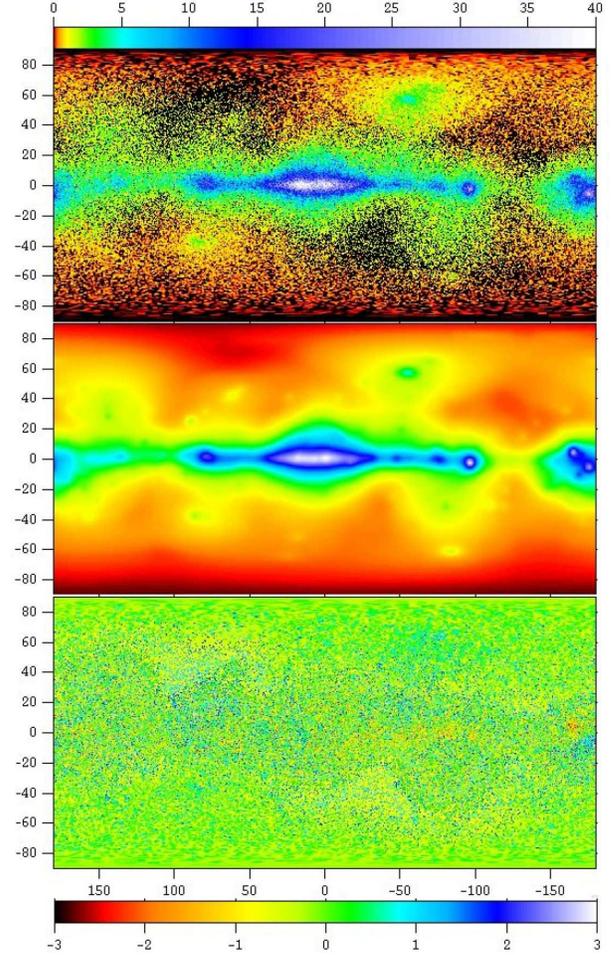}
\caption{From top to bottom: EGRET photon map, best-fit photon map and residuals. The first two maps are expressed in photons/bin while the latter is in $\sigma$ units. All maps are given in Galactic coordinates for the 70 to 150 MeV energy band. The data and model maps are displayed in logarithmic scale, the residuals in linear scale.}
\label{fig:counts}
\end{center}
\end{figure}

\begin{figure}
\begin{center}
\includegraphics[width=8cm]{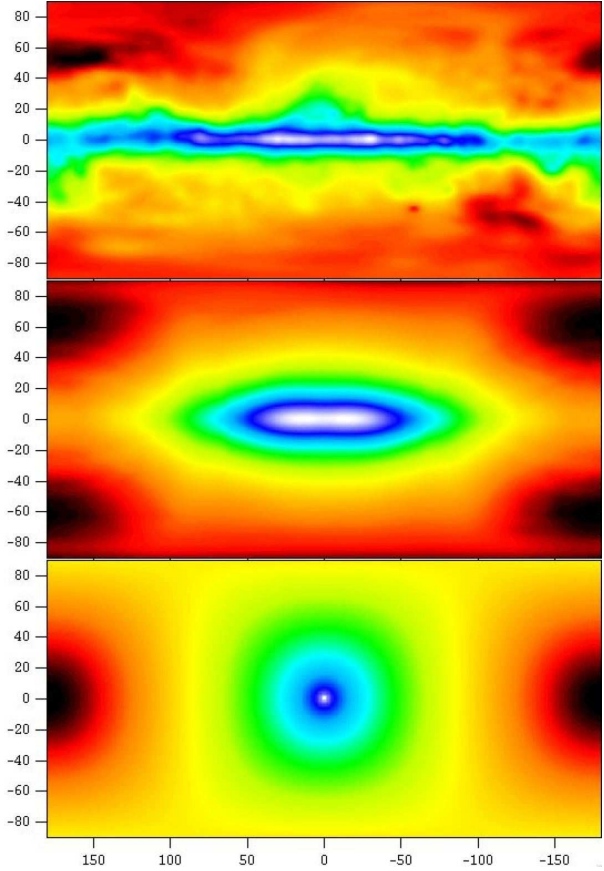}
\caption{Predicted photon counts in the merged 70 to 150 MeV energy band originating from $\pi^{0}$-decay and bremsstrahlung radiation in the $HI$ and $H_{2}$ gas (top), inverse Compton scattering of the interstellar radiation field (middle), and PBH emission from the Moore dark matter distribution (bottom). All maps are displayed in Galactic coordinates with different logarithmic scales to easily compare the radial profiles.}
\label{fig:intensity}
\end{center}
\end{figure}

\begin{figure}
\begin{center}
\includegraphics[width=8cm]{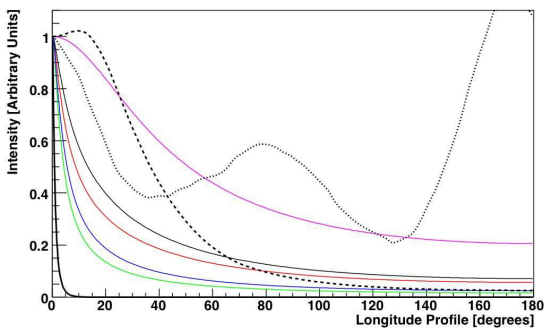}
\caption{Normalized longitude profile of the EGRET PSF (black continuous thick), exposure (black dotted) and inverse-Compton (IC, black dashed) maps, as well as the profiles of the PBH emission maps obtained for the dark matter distributions of Moore (red), Moore$_\mathrm{c}$ (green), NFW (black thin), NFW$_\mathrm{c}$ (blue), and isothermal (magenta). All profiles are given for energies $E > 100$~MeV and are centered on b = 0$^{\circ}$. The IC and PBH profiles include the convolution by the PSF at these energies.}
\label{fig:profiles}
\end{center}
\end{figure}

Figure~\ref{fig:counts} shows the EGRET count map in the merged 70 to 150 MeV band, together with the predicted count map obtained for the same energy range from the best-fit model without any PBH contribution. The bottom map gives the residuals expressed in $\sigma$ units, $(N_{obs}(l,b)-N_{pred}(l,b))/N_{pred}(l,b)^{1/2}$. The latter is smooth and shows no obvious positive residual toward the inner Galaxy, so the combination of interstellar, isotropic, and point-source components reasonably fit the observed $\gamma$-rays. The patches of small, but systematic, residuals spanning angular scales of 20 to 30 degrees in various places across the sky are correlated with the exposure map and are due to off-axis aberrations and systematic uncertainties in the off-axis effective area of the telescope.

A useful constraint on the PBH number density can be derived if their cumulative emission in the Milky Way is well separated in space and energy from the main Galactic and isotropic contributions. The prominent lens-like source of $\gamma$-rays seen in the data and model maps at the centre of Figure \ref{fig:counts} results from the enhanced cosmic-ray density and interstellar radiation field in the inner Galaxy. The photon field includes the stellar light from the Galactic bulge which is concentrated within about 10 degrees of the Galactic centre (Moskalenko et al. \cite{moskalenko06}). Because dark matter piles up near the centre as well, one expects some degeneracy between the integrated intensity from PBHs and the inverse-Compton radiation, but they exhibit different ellipticities (latitude to longitude ratios) that allow their separation in the fit (see Figure \ref{fig:intensity}). The intensities in Figure \ref{fig:intensity} are integrated from 70 to 150 MeV where the PBH spectrum peaks, but where the EGRET angular resolution is poorer. It illustrates that the two maps still differ after convolution with the wide instrument function. Figure~\ref{fig:profiles} shows the relative longitude profiles, after convolution with the PSF, of the inverse-Compton component and PBH emission for various dark matter distributions. All PBH profiles but the isothermal one have smaller angular scales than the inverse-Compton one. The rapid drop of the isothermal distribution outside 30 degrees, together with a different spectrum, reduces the degeneracy between the two components. The PSF profile, which is also displayed in Figure \ref{fig:intensity}, appears to be significantly sharper than even the most concentrated dark matter distributions, thus allowing separation of the PBH contribution from a soft point-source. 

The likelihood probability can serve to test the presence or absence of PBH emission in addition to the other radiative components. We derived the likelihood ratio test statistic defined as $TS = 2 [\ln(L_{1}) - \ln(L_{0})]$, where the likelihood values $L_{1}$ and $L_{0}$ are optimized respectively with and without a PBH contribution in equation \ref{eq:eqRing}. Asymptotically, the $TS$ distribution follows a  $\chi^{2}_1$ distribution and the detection significance of PBH emission is $\sqrt{TS} \, \sigma$.

\begin{figure}
\begin{center}
\includegraphics[width=8cm]{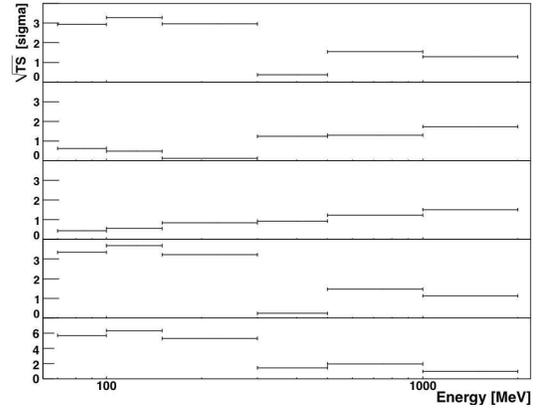}
\caption{Comparison of Galactic emission model, with and without the PBH contribution, to the EGRET count map in six energy bands. The result expressed in $\sqrt{TS}$, or detection significance in unit of $\sigma$, is shown for five dark matter distributions: Moore, Moore$_\mathrm{c}$, NFW, NFW$_\mathrm{c}$ and isothermal (from top to bottom).}
\label{fig:sigma}
\end{center}
\end{figure}

The $\sqrt{TS}$ results obtained in the six energy bands are given in Figure \ref{fig:sigma}. They yield no significant detection at the 3 $\sigma$ level of any PBH emission for the dark matter distributions of Moore, Moore$_\mathrm{c}$, NFW, and NFW$_\mathrm{c}$  distributions. A small likelihood increase, amounting to 6 $\sigma$, is found for the isothermal distribution at low energy. It vanishes above 300 MeV. Although a low-energy excess is expected from black hole emission, the signal is not significant enough to be interpreted as a signature of PBH activity.

The main reason is that the intensity does not decrease in the 150-300~MeV energy range as it should (it should amount to only 7.5\% of the total PBH intensity). We also point out that the isothermal distribution has the nearest angular profile to the inverse-Compton one. The additional emission that is attributed by the fit to the isothermal map is likely to compensate for a lack of inverse-Compton emission in the central part of the Galaxy. One should then note that gas in the dense central molecular zone must be mapped using rarer CO isotopes which are not part of the present gas maps. The central molecular zone (CMZ, about $500 \times 500 \times 50$~pc) has a high average hydrogen density ($n_H \approx 50\, \mathrm{to}\,100\,\mathrm{cm}^{-3}$, Launhardt et al \cite{launhardt02}, Ferri\`ere \cite{ferriere07}), about 90\% of the interstellar gas being in the form of dense clouds ($10^3\,\mathrm{cm}^{-3}$) with higher values in cloud cores. The remaining 10\% would form a diffuse homogeneous intercloud medium easily pervaded by GeV cosmic rays (which produce most of the 100~MeV photons). The Galactic centre region also abounds in massive stars, in particular in young and dramatically active OB associations (the Arches, Quintuplet, and Central clusters, Figer et al. \cite{figer00, figer08}) where SNII shocks and colliding stellar winds from densely packed Wolf-Rayet and O stars are favorable to cosmic-ray acceleration. So, we can expect both a strong enhancement in the $\gamma$-ray intensity corresponding to CMZ and its surroundings when cosmic rays diffuse out, and a different spectrum from freshly accelerated cosmic rays compared to the local one. Both these effects may have been observed at TeV energies and in the radio (see Grenier \cite{grenier08} for a review), but they have not been included in the present model. Furthermore, the 2$\sigma$ likelihood increase found above 1~GeV for all PBH distributions may be related to the present use of the Earth in-situ cosmic-ray electron spectrum to derive the inverse-Compton intensity, whereas the EGRET data requires a harder IC spectrum to fit the data. This GeV excess, probably of instrumental origin (Stecker et al. \cite{stecker08}), is not taken into account in the present model. Given these significant uncertainties in the actual shape and spectrum of the inverse-Compton emission in the inner Galaxy and from the presence of an additional contribution from the CMZ gas, we do not consider the likelihood increase related to the isothermal PBH component as significant.

To obtain a fair upper limit on PBH emission, we merged the 70-100~MeV and 100-150~MeV energy bands where 72\% of the signal is expected. We used the likelihood test ratio to increase the PBH contribution until the Log-likelihood drops by 9 from the maximum value obtained under the null hypothesis (without PBH emission). As required by the definition of TS, all other free parameters (interstellar, isotropic, and sources) were optimized as we changed the PBH intensity. In other words, we increased the PBH contribution until the predicted map was so degraded that there was a $10^{-3}$ chance that random fluctuations from the null+PBH model yield as good a fit to the data as the null model. The derived upper limits are presented in Table \ref{tab:results}.

\section{Results}
\label{sec:results}

We have determined the maximum fractions, $f$, of dark matter in the form of PBHs in our Galaxy which decay presently and participate to the Galactic $\gamma$-ray emission for different possible dark matter distributions (see Table~\ref{tab:results}). Knowing the dark matter density in the Universe, $\Omega_{DM} = 0.228\pm 0.013$ (Komatsu et al. \cite{WMAP5}), these fractions can be translated into $\Omega_{PBH}(M_{\star})$, the cosmological density of PBHs with mass equal to $M_{\star}$: $\Omega_{PBH}(M_{\star}) \approx f \Omega_{DM}$. We get values ranging from $2.4\,10^{-10}$ to $2.65\,10^{-9}$ which leads to a significant improvement for the concentrated Moore$_c$ and NFW$_c$ distributions over previous bounds deduced from the extragalactic $\gamma$-ray background, i.e. $10^{-8}$ (Carr \& McGibbon \cite{carr98}) and  $3\,10^{-9}$ (Barrau et al. \cite{barrau03}), and from the number of antiprotons, $3\,\mathrm{to}\,4\,10^{-9}$ (Maurin et al. \cite{maurin01}, Barrau et al. \cite{barrau02b}). It is worth noting that our method is independent of the poorly known initial mass function of PBHs. Our results are consistent with a completely independent constraint derived from the apparent absence  of gravitational capture of PBHs with $M > M_\mathrm{\star}$ by solar-type stars and white dwarfs (Roncaldi et al. \cite{roncaldi09}).

\begin{table}[htdp]
  \caption{Constraints on PBHs of mass  $M_{\star}$ for different dark matter distributions$^{\mathrm{a}}$.}
  \begin{tabular}{|l|l|c|c|}
\hline
DM distribution   & $f(M_{\star})$  & $\Omega_{PBH}(M_{\star})$ & $\beta(M_\star)$  \\
\hline
Moore        & $ 6.04 \pm 0.05\,10^{-9}$ & $1.38\,10^{-9}$ & $0.98\,10^{-27}$  \\
Moore$_\mathrm{c}$      & $ 1.07 \pm 0.07\,10^{-9}$  & $0.24\,10^{-9}$ & $0.17\,10^{-27}$  \\
NFW           & $ 6.70 \pm 0.05\,10^{-9}$  & $1.53\,10^{-9}$ & $1.08\,10^{-27}$  \\
NFW$_\mathrm{c}$         & $ 1.93 \pm 0.08\,10^{-9}$  & $0.44\,10^{-9}$ & $0.31\,10^{-27}$ \\
isothermal & $11.62 \pm 0.04\,10^{-9}$ & $2.65\,10^{-9}$ & $1.87\,10^{-27}$  \\
\hline
  \end{tabular}

\begin{list}{}{}
\item[$^{\mathrm{a}}$] $f$ is the maximum fraction of dark matter in the form of PBHs, $\Omega_{PBH}(M_{\star})$ is the corresponding cosmological density, and $\beta$ is the fraction of regions undergoing collapse when the mass enclosed within the cosmological horizon was $M_{\star}$.
\end{list}


  \label{tab:results}
\end{table}

\section{Astrophysical consequences}
\label{sec:consequences}

\subsection{Local PBH density}
We can derive directly the local density of PBH of mass $M_\star$ (and the mean distance between them) from the dark matter density in the solar vicinity (i.e. $\rho_\mathrm{DM}(\odot) \approx 0.3$~GeV/cm$^3$). We get values ranging from $3.2\,10^{-10}$ to $2.0\,10^{-9}\, \mathrm{GeV}\,\mathrm{cm}^{-3}$ corresponding to a maximum number density of $M_\star$ PBHs within the range $3.3\,10^7$ to $2.1\,10^8$  per pc$^3$. These values, in turn, can be used to derive an upper limit of the present explosion rate of PBHs (Schroedter et al. \cite{schroedter08}, Petkov \cite{petkov08} and references therein).

To estimate the local explosion rate of PBHs, we rely on the Wright's pioneering analysis of the
$\gamma$-ray PBH emission of the dark halo (Wright \cite{wright96}). This author deduces that the concentration factor of PBH, which is the ratio of the local to the cosmic PBH densities should be less than $2 - 12\,h^{-1}\,10^5$~pc$^{-3}$, i.e.$2.74\,10^5 - 1.64\,10^6$~pc$^{-3}$  taking the reduced Hubble constant $h = 0.73$. Combining this estimate with the work of Halzen et al. (\cite{halzen91}), Wright deduces a limit on the explosion rate $< 0.07 - 0.42$~pc$^{-3}$~yr$^{-1}$. Since our concentration factor, is $\approx 2.3\,10^5$ (because $\rho_\mathrm{DM}(\odot) \approx 0.3$~GeV/cm$^3$ while $\rho_\mathrm{DM} \approx 0.23\,\rho_c \approx 1.3\,10^{-6}$~GeV~cm$^{-3}$, where $\rho_c$ is the critical density),  by simple scaling we get an explosion rate $0.059$~pc$^{-3}$~yr$^{-1}$.

This limit should be compared to the observational ones. However the behaviour of high temperature PBH (which has low mass) is poorly known, and different models of the final burst can give vastly different detection limit. The final fate of PBH and the ability to observe their final decay depends on the model of particle physics chosen to describe their high temperature behaviour. In the standard model the number of elementary particle species never exceeds 100, and the likelihood of detecting such explosions  is very low. However, the physics of the QCD phase transition around 157-180 MeV is still uncertain. The prospect of detecting PBH explosions would be improved in less conventional particle physics models. For instance in the Hagedorn model the number of particle species, including resonances, proliferates at the QCD temperature. The resulting burst of radiation is therefore much shorter and intense than in the standard case. In the standard model case, the limit derived here on  the local rate density of exploding PBH provides a very difficult target for all techniques to direct detection of PBH explosions through $\gamma$-rays. By comparison, our explosion rate is:
\begin{itemize}
  \item near the EGRET observational limit (0.05~pc$^{-3}$~yr$^{-1}$, see Fichtel et al. \cite{fichtel94}), assuming that the last $6\,10^{13}$~g  of PBH rest mass evaporates, producing $10^{34}$~ergs of 100~MeV $\gamma$-rays in less than a microsecond, based on the Hagedorn model of high energy, which remains the most stringent $\gamma$-ray observational limit at present time;
  
  \item higher than the Fermi expected detection rate ($\approx 0.02$~pc$^{-3}$~yr$^{-1}$ for 2 years observations,  see Schroedter et al. \cite{schroedter08});
  
  \item higher than the Maki et al. (\cite{maki96}) antiproton limit  (0.017~pc$^{-3}$~yr$^{-1}$) but  independent of the delicate cosmic ray diffusion coefficient, since $\gamma$-rays travel in straight lines.
\end{itemize}

\subsection{Multidimensionals PBHs}

Interestingly enough, extending this study to higher dimensions (Argyres et al. \cite{argyres98}), we find that the temperature of the PBH in a 11D  spacetime, with a lifetime equal to that of the Universe, is close to that of 4D ones. Indeed their Schwarzschild radius (5~fm) is very close to the radius of the extradimensions (7~fm) if the fundamental Planck mass is of order 1~TeV, as suggested by Arkani-Hamed, Dimopoulos and Dvali (\cite{arkani98}). Consequently, they are expected to behave like ``normal" 4D black holes. For such a high number of extra dimensions, the Kaluza-Klein modes of the graviton are so energetic (being proportional to the inverse radius of the extra dimensions) that they are not excited at the PBH temperature. Thus the branching ratios of a 4D black hole can be applied to them safely. Their calculated temperature being 24 MeV they are almost undistinguishable from 4D PBH. More work is necessary to settle the question since when the black hole radius is of the order of the Planck length we enter in the domain of pure quantum gravity, poorly understood. Thus, qualitatively at least, the conclusions of this work could apply to superstring black holes. We will discuss this issue in much more details in a forthcoming paper.

\subsection{Cosmological consequences}

PBHs should be the first objects to form by gravitational collapse of density inhomogeneities in the early Universe, and as such they are a unique tool to probe the very small cosmological scales, an order of magnitude smaller than that explored by measurements of the cosmic microwave background and/or large scale structures (Carr \& MacGibbon \cite{carr98},  Khlopov \cite{khlopov08}, Barrau et al. \cite{barrau03}). The fraction of regions undergoing collapse when the mass enclosed within the cosmological horizon is $M$ is determined by the equation of state of the medium in which PBHs form ($P = \gamma \rho$) and  the root mean square amplitude $\epsilon$ of the fluctuations entering the horizon at that epoch (Carr \& MacGibbon \cite{carr98}) through 
$$\beta(M) = \epsilon(M)\,\exp(-\gamma^2/2\epsilon(M)^2),$$
provided the fluctuations are gaussian and spherically symmetric and PBH production takes place in the radiation dominated era ($\gamma = 1/3$). Thus $\beta(M)$ offers a link between $\Omega_{PBH}(M)$ and $\epsilon(M)$. Numerically, 
 $$\Omega_{PBH}(M) \approx 10^{18} \beta(M)\,\left( \frac{M}{10^{15}\,\mathrm{g}} \right)^{-1/2}.$$
Several constraints on $\beta$ have been derived in different mass ranges from primordial nucleosynthesis and cosmic microwave background, i.e. for PBHs of mass smaller than $M_{\star}$ (see Figure~1 of Carr \& MacGibbon \cite{carr98}). Our values of $\Omega_{PBH}(M_{\star})$ translates to $\beta(M_{\star})$ values listed in Table~\ref{tab:results}.  Blais et al. (\cite{blais03}) propose a different $\Omega$-$\beta$ relation based on a model in which the power spectrum at small scale is boosted and where
$$\Omega_{PBH}(M)\, h^2 = 6.35\,10^{16}\,\beta(M)\, \left( \frac{M}{10^{15}\,\mathrm{g}} \right)^{-1/2}.$$
Our upper limit associated with $h = 0.7$ (Komatsu et al. \cite{WMAP5}) leads to $\beta(M_{\star})$ values around 10 times smaller than in the previous case.  One natural source of fluctuations would be inflation and, in this context, $\epsilon(M)$ depends implicitly on the inflation potential, but this subject is beyond the scope of the present work.

\section{Conclusion and perspectives}

Our constraint on the maximum density of primordial black holes is more stringent, or at least comparable, than the previous ones. The method proposed is more direct and robust due to the fact that Galactic $\gamma$-ray astronomy around 100 MeV has reached a level of unprecedented quality both from the theoretical and observational points of view.

On the observational side, the large improvement in both sensitivity and angular resolution of the new generation Large Area Telescope (LAT) on board the Fermi Gamma-ray Space Telescope will allow to strengthen the upper limit on $\Omega_{PBH}(M_{\star})$. The central regions of the Galaxy will remain difficult to model precisely, but the sharper PSF will help separate low-energy PBH emission from the wider scale of the inverse-Compton radiation and from a population of point-sources in the Galactic bulge. The low-energy threshold of the LAT (20 MeV), the large photon statistics and very smooth exposure resulting from the continuous all-sky survey currently underway will help take advantage of the unusual spectrum of PBH emission to separate it from the harder and power-law spectra of inverse-Compton and gas emission, or from the more isotropic contamination from soft instrumental backgrounds. The low statistics of EGRET has not permitted to fully exploit this aspect. 

On the theoretical side, the extension of this method to non standard light particles would shed new light on particle physics. Indeed, a PBH radiates any and all particles with rest masses which are substantially less than its current temperature, including graviton and hypothetical light dark matter particles (Boehm et al. \cite{boehm04}, Fayet \cite{fayet04}), axions and  light scalars. Concerning light fermionic dark matter invoked to explain the low-energy positron excess deduced from the 511~keV line observation of the inner Galaxy (Boehm et al. \cite{boehm04}, Fayet \cite{fayet04}), it would correspond to introducing an additional neutrino species which, opening a new decay chain, would slightly lower the $\gamma$-ray yield and alter the lifetime of PBHs. The effect of non-standard particles on the black hole radiation is postponed to a forthcoming paper.

\begin{acknowledgements}
We are grateful to Seth Digel for producing the HI and CO rings map.  We also thank Andrew Strong and Igor Moskalenko for providing access to the GALPROP package.
\end{acknowledgements}

\end{document}